\documentclass[aps,prc,preprint,showpacs,showkeys,floatfix,nofootinbib,%
superscriptaddress]{revtex4}
\usepackage{graphicx}
\usepackage{multirow}
\usepackage{bm}

\begin{document}

\title{Bridging over $p$-wave $\pi$-production and weak processes in
few-nucleon systems with chiral perturbation theory
}

\author{Satoshi X. Nakamura}\email{snakamura@triumf.ca}
\affiliation{Theory Group, TRIUMF,
4004 Wesbrook Mall, Vancouver, BC V6T 2A3, Canada}

\begin{abstract}
We focus on 
a powerful aspect of chiral perturbation theory ($\chi$PT) which
provides us with a ``bridge'' over different reactions.
For this purpose,
we study a contact operator which contributes $p$-wave
$\pi$-production and several weak processes.
We fix the unknown coupling of the contact operator
 using a matrix element of a low-energy weak
 process ($pp\to de^+\nu_e$), and
then calculate the partial wave amplitude ($a_0$) for the $p$-wave
$\pi$-production ($pp\to pn\pi^+$).
We find that the chiral operator including the contact term does not
 reproduce $a_0$ extracted from data,
showing that the bridging over reactions with significantly
 different kinematics is not necessarily successful.
We argue the importance of a specific higher order calculation.
In order to gain an insight into a higher order calculation,
we consider a higher order counter term, and find that
the energy dependence of $a_0$
is then consistent with the data.
\end{abstract}

\preprint{TRI-PP-07-27}

\maketitle

\section{Introduction}

Since the beginning of 1990's, chiral perturbation theory ($\chi$PT) has
been extensively applied to few-nucleon system.
An advantageous point inherent in $\chi$PT is that it bridges over
different reactions in a model-independent manner.
Once couplings (the so-called low-energy constant, LEC) included in a
chiral operator are fixed using data for one of the reactions, then the
other reactions are predicted using the same operator.
An interesting interaction in this context is,
 ${\cal L} = \tilde{d} N^\dagger S \cdot u N N^\dagger N$ ,
with
$ f_\pi u_\mu = -\tau_a \partial_\mu \pi_a 
- \epsilon_{3 ab} V_\mu \pi_a \tau_b 
+ f_\pi A_\mu + \cdots $. 
The spin operator is $S$,
and the external vector (axial) current is $V_\mu$ ($A_\mu$).
The constant $\tilde{d}$ is a LEC.
This contact interaction
contributes to 
the three-nucleon force, and several processes such as
the $p$-wave $\pi$-production ($pp\to pn\pi^+$\cite{hanhart_p-pi}),
and weak processes
(tritium $\beta$-decay, $pp\to de^+\nu_e$\cite{park}).
We can fix $\tilde{d}$ using one of the above reactions,
and predict the others.

Such a ``bridging program'' has been done in several works.
One of them was done by Park {\it et al.}~\cite{park}, where they
fixed $\tilde{d}$ using the experimental tritium $\beta$-decay rate, and
performed a parameter-free calculation of the weak proton capture by a
proton (or $^3$He).
Another work was due to Hanhart {\it et al.}~\cite{hanhart_p-pi}.
The authors calculated the partial wave amplitude ($a_0$) for the
$p$-wave $\pi$-production ($pp\to pn\pi^+$).
Even though this work was not a fully consistent bridging program,
they showed that
the use of $\tilde{d}$ fixed by three-nucleon observables
consistently reproduces $a_0$
extracted from data~\cite{flammang}.

In this work (see Ref.~\cite{sxn} for a full account), we
investigate more seriously how reliably
the bridging program, an important aspect of $\chi$PT, works.
We believe that our investigation is important because we have
sometimes seen an argument which supposes, without a serious
test, that the bridging program works.
For this purpose, we calculate the partial wave amplitude ($a_0$) for the
$p$-wave $\pi$-production ($pp\to pn\pi^+$), 
with $\tilde{d}$ fixed by a low-energy weak process.
This obviously provides a stringent test of $\chi$PT,
because the two reactions are strong and weak processes, and are 
quite different in kinematics.

\section{Chiral $p$-wave $\pi$-production operator}

We use the following $\pi$-production operators by referring to
Ref.~\cite{hanhart_p-pi} in which
the operators were derived using a counting
rule based on an expansion parameter, $\sqrt{m_\pi/m_N}$;
the nucleon (pion) mass is denoted by $m_N$ ($m_\pi$).
The leading order (LO, ${\cal O}(1)$) operator is the one-body direct
production of the pion off the nucleon.
Another LO mechanism we consider is the $\Delta$-excitation followed
by the $\pi$ emission.
In Ref.~\cite{hanhart_p-pi}, the authors used the wave function which
explicitly includes the $\Delta$ component, and considered the one-body
operator which produces the pion with the $\Delta$ deexcited to the
nucleon.
Because we use nuclear wave functions with only the nucleonic degrees of
freedom, we alternatively use a two-body operator in which the $\Delta$
is excited either by the $\pi$-exchange or by
a contact interaction.
Next we discuss next-to-leading order (NLO, ${\cal O}(m_\pi/m_N)$) terms:
the recoil correction to the LO terms;
a pion rescattering through
the vertices whose strength are $c_3$ and $c_4$,
or through
the Weinberg-Tomozawa term or its Galilean correction;
a pion emission from the contact term 
whose coupling constant is $\tilde{d}$.
The unknown coupling, $\tilde{d}$, will be determined in the
next paragraph.

We start with a benchmark calculation.
We employ the same axial current operator used in Ref.~\cite{park}.
The $\tilde{d}$ value in this operator has been fixed using the tritium
$\beta$-decay rate.
We calculate the Gamow-Teller matrix element for the
low-energy $pp\to de^+\nu_e$ reaction and
regard it as the benchmark.
Our axial current is different from Ref.~\cite{park}
in that we consider the $\Delta$ explicitly.
We calculate the matrix element using our operator and
fit $\tilde{d}$ to the benchmark result.
We use several combinations of the $\pi N\Delta$ coupling ($h_A$), the $NN$
potential and the cutoff ($\Lambda$), and obtain the corresponding $\tilde{d}$
values with a natural strength.

\section{Result}
\begin{figure}
\begin{minipage}[t]{67mm}
\begin{center}
\includegraphics[width=65mm]{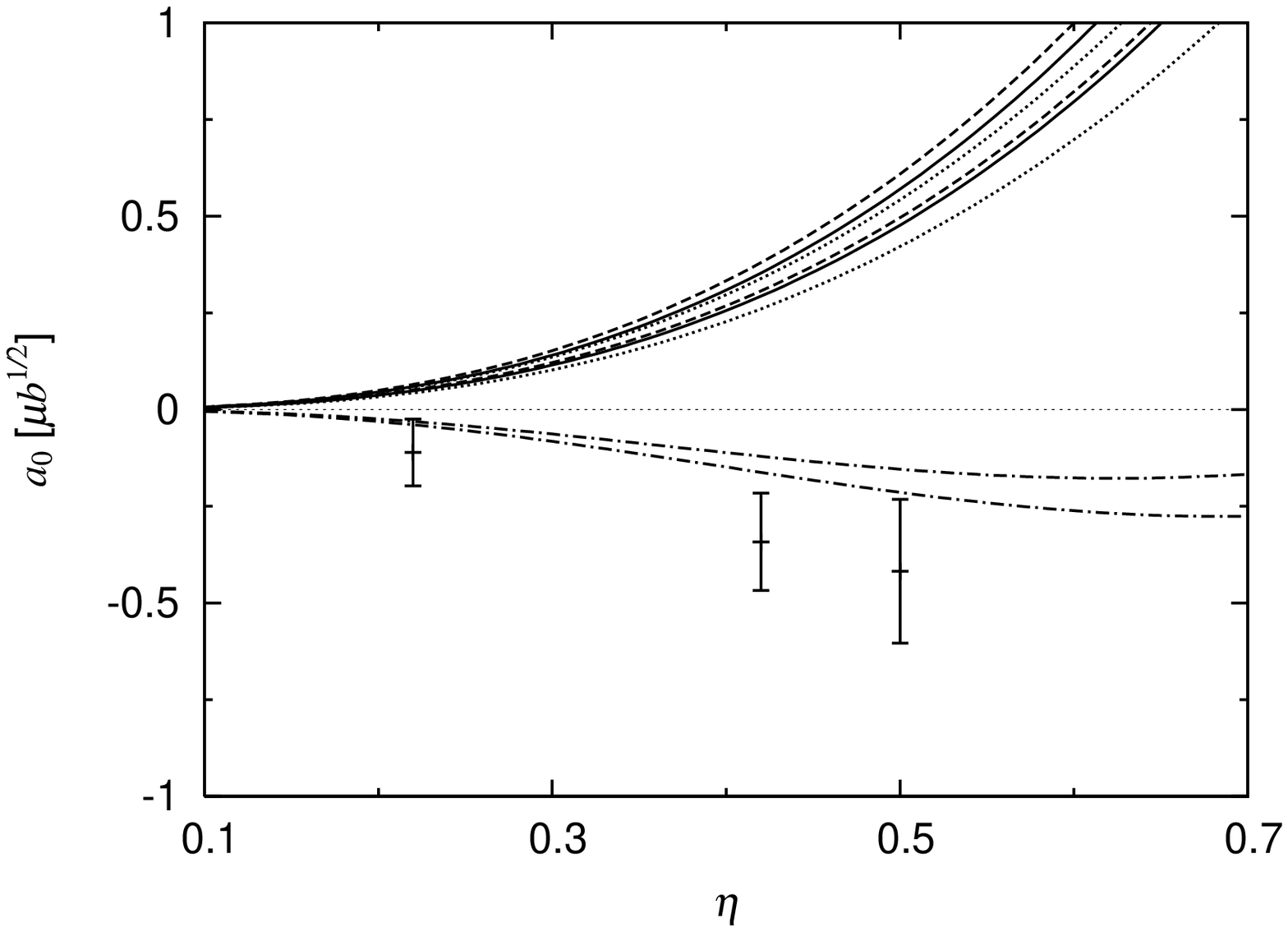}
\caption{The amplitude $a_0$ for $pp\to pn\pi^+$.
The upper (lower) solid, dashed and dotted curves correspond to
$\Lambda =$ 500, 600 and 800~MeV, respectively, $h_A =$ 2.10 (2.68).
The lower (upper) dash-dotted curve is obtained with $\tilde{d} =$ 0,
$\Lambda =$ 800~MeV, $h_A =$ 2.10 (2.68).
Data are from Ref.~\cite{flammang}.}
\label{fig1}
\end{center}
\end{minipage}
\hspace{2mm}
\begin{minipage}[t]{65mm}
\begin{center}
\includegraphics[width=65mm]{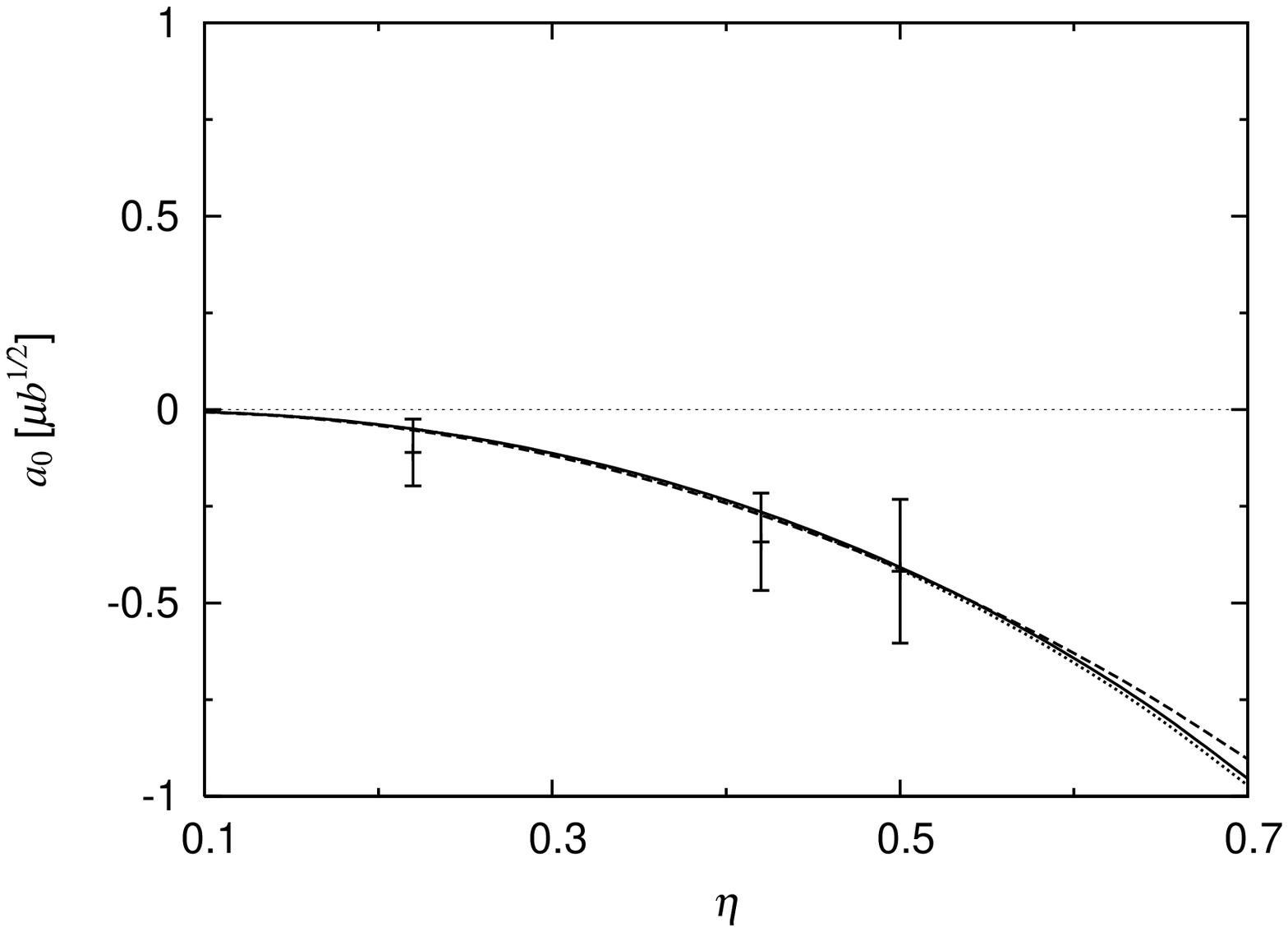}
\caption{\label{fig2}
The amplitude $a_0$ for $pp\to pn\pi^+$.
The chiral NLO $\pi$ production operator 
plus the $\tilde{e}$-term is used.
The solid, dashed and dotted curves correspond to
$\Lambda =$ 500, 600 and 800~MeV, respectively, and $h_A =$ 2.10.
The other features are the same as Fig.~\ref{fig1}.}
\end{center}
\end{minipage}
\end{figure}

With the chiral operator fixed in the previous section,
we calculate $a_0$ for $pp\to pn\pi^+$.
In Fig.~\ref{fig1}, we present 
$a_0$ as a function of $\eta \equiv q_\pi^{max}/m_\pi$, where
$q_\pi^{max}$ is the maximum pion momentum.
We used the CD-Bonn $NN$ potential and two (three)
choices of $h_A$ ($\Lambda$).
Our result is rather different from the data.
For comparison, we show $a_0$ obtained without the $\tilde{d}$
term.
The inclusion of the $\tilde{d}$ term makes the disagreement worse;
even the sign of $\tilde{d}$ fixed by the low-energy weak
process is inconsistent with the data.
We change the values of $\Lambda$, $h_A$ and the $NN$ potential,
however, the situation of the disagreement does not change.
This result shows that the bridging program among reactions with significantly
different kinematics is not necessarily successful.

The failure of the bridging program is understandable
if we recall the success of the chiral nuclear force which
describes the $NN$ scattering in a wide energy region.
This is partly because the LECs have been fitted to data from the
same energy region. 
In order to accurately describe the two reactions in different
energy regions, data from the both energy region are necessary to fix
the LECs.
It is also expected that higher order terms are necessary to accurately
reproduce the data from the wide energy region.

To explore, even roughly, a higher order calculation,
we perform a simple extension of the previous calculation by
adding 
${\cal L}_{\rm CT}^{(2)} 
        =        \tilde{e}
        N^{\dagger}{\boldsymbol\tau}\vec{\sigma}\cdot\vec{\nabla}{\boldsymbol\pi} N\,
        (N^{\dagger} \vec{\nabla}^2N + {\rm h.c.})$
to the NLO operator.
Now we have the two independent LECs: $\tilde{d}$ and $\tilde{e}$.
We fix these two LECs to reproduce two quantities:
the benchmark result for the matrix element of
$pp\to de^+\nu_e$;
$a_0$ for $pp\to pn\pi^+$ at $\eta = 0.5$ from data.
We obtain a set of $\tilde{d}$ and $\tilde{e}$
with a natural strength in this way.
We calculate $a_0$ with the extended operator and show the result in
Fig.~\ref{fig2}.
By construction, the calculated $a_0$ goes through the central
value of the data at $\eta =$ 0.5.
We examined the dependence of $a_0$ on $h_A$, $\Lambda$ and the $NN$
potential, and found that 
the $\eta$ dependence of $a_0$ is consistent with the data.

\section{Summary}

We determined $\tilde{d}$, the LEC of the contact term, using the
weak process and then used it to predict 
$a_0$ for $pp\to pn\pi^+$.
Through this work, we tried to explore the power of $\chi$PT
that enables us to bridge different reactions.
Our prediction of $a_0$ with the NLO operator does not agree
with the data.
The result indicates that the bridging program between 
reactions with significantly different kinematics is not always successful.

\section*{Acknowledgments}
The Natural Sciences and Engineering Research Council of Canada is
thanked for financial support. TRIUMF receives federal funding via a
contribution agreement through the National Research Council of Canada.


%
\end{document}